\documentstyle[aps]{revtex}
\begin{document}
\title{Quantum measurement of coherence in coupled quantum dots}
\author{H.M. Wiseman~${}^{1}$, Dian Wahyu Utami~${}^2$, He Bi Sun~${}^2$, G.J.
Milburn~${}^2$, B.~E.~Kane~${}^3$,A.Dzurak~${}^4$, R.G.Clark~${}^4$}
\address{
${}^1$School of Science, Griffith University, Qld. 4111 Australia\\
${}^2$Centre for Quantum Computer Technology ,The University of Queensland, Qld. 4072 Australia\\
${}^3$Laboratory for Physical Sciences, College Park, Maryland 20740\\
${}^4$Centre for Quantum Computer Technology, The University of New South Wales, Sydney 2052, Australia.}
\date{\today}
\maketitle

\begin{abstract}
We describe the conditional and unconditional dynamics of two coupled
quantum dots when one dot is subjected to a measurement of its occupation
number using a single electron transistor (SET). The measurement is made
when the bare tunneling rate though the SET is changed by the occupation
number of one of the dots. We show that there is a difference between the
time scale for the measurement-induced decoherence between the localized
states of the dots and the time scale on which the system becomes localized
due to the measurement. A comparison between theory and current experiments
is made.
\end{abstract}


%\draft

%\newpage

%\begin{multicols}{2}

\section{Introduction}

There have recently been a number of suggestions for a quantum computer
architecture that use quantum dots of varying kinds\cite
{Kane98,Loss98,Schoen97}. If these schemes are to be practical many
important physical questions need to be answered, one of which is how to
readout physical properties such as charge or spin at a single electron level%
\cite{Gurvitz97,Schoen98}. In this paper we present a quantum trajectory
analysis of a general scheme to readout a single electronic qubit using a
single electron transistor (SET). We adopt a general phenomenological
description of the SET in which the tunneling rate through the SET is
conditioned on the occupation or otherwise of a nearby quantum dot.

We consider two spatially separated quantum dots which are strongly coupled
so that delocalized states of their relevant degrees of freedom can form. To
be specific, we imagine each dot to have a single electronic bound state
that can be occupied. Thus the average occupation number of each dot must be
less than unity. This restriction can easily be removed to account for spin,
or multiple electron states. We label each dot with an index $1,2$ and let $%
c_i,c_i^\dagger$ represent the Fermi annihilation and creation operators for
each single electron state (see figure 1).

The two dots are strongly coupled via the tunnel coupling Hamilton
\begin{equation}
V=i\hbar\frac{\Omega}{2}(c_1^\dagger c_2-c_2^\dagger c_1)  \label{tunnel}
\end{equation}
Thus the total Hamiltonian of the two-dot system is
\begin{equation}
H=\hbar\sum_{i=1}^2 \omega_ic_i^\dagger c_i+V
\end{equation}
In what follows we will work in an interaction picture and assume that the
energies of each bound state are equal (again this can be relaxed). Coulomb
blockade effects have been ignored at this stage, but can easily be included
without significantly changing the results of this paper.

The single particle eigenstates of this Hamiltonian are even and odd
superpositions of the bare states of each well. Such states are thus
delocalized over the two-dot system and are sometimes called 'molecular
states' in the literature\cite{Blick98}. The localized states can then be
represented as an even and odd superposition of the delocalized states. The
localized states are not stationary; rather, the system will periodically
oscillate between them. That is to say the system will tunnel coherently
between the two dots.

To this coherent system we add a measurement device which  determines the
presence of an electron on one of the dots, say dot $1$, which we shall
refer to as the target (see figure \ref{fig1}). The model is based on a SET
tunnel junction containing a single bound state on the island. The
interaction between the target and the SET is via a Coulomb blockade. Thus
the interaction Hamiltonian between the SET and the target must commute with
the target electron number operator , $c_1^\dagger c_1$. This makes it a QND
(quantum nondemolition measurement) of electron number\cite{WallsMilb94}.
The Coulomb blockade changes the current flowing through the SET. In simple
terms if there is no electron on the target the island state is biased so as
to allow little or no current to flow though the SET. This is the quiescent
state of the SET. However when there is an electron on the target, the
coulomb blockade shifts the bound state on the island to allow a greater
current to flow through the device (see figure \ref{fig2})

We derive a master equation to describe the behaviour of the target system.
This master equation describes the unconditional evolution of the measured
system when the results of all measurement records (that is current records)
are averaged over. This will tell us the rate at which coherence in the
target system is destroyed by the measurement. However we also need to know
how the system state depends on the actual current through the device in
order to determine how quickly the conditional state of the electron becomes
localized which is measure of the quality of the measurement.

One approach to this problem is to keep track of many different states of
the system, corresponding to the different numbers of electrons which have
tunneled through the SET. This is the approach used for example in Ref.~\cite
{Sch99}. Here we adopt an alternate method which gives a more intuitive
picture for the conditional dynamics. We use a conditional {\em stochastic}
master equation which gives the evolution of the measured system,
conditioned on a particular realization of the measured current. The
instantaneous state of the target conditions the measured current while the
measured current itself conditions the future evolution of the measured
system in a self consistent manner. This approach to measurements has been
variously called the quantum trajectory method \cite{Carmichael93} or
quantum monte carlo method \cite{DalCasMol92}.

\section{SET Model}

Consider a two dot system with the coupling in Eq.~(\ref{tunnel}). If the
electron is in dot-2 the quiescent rate of current tunneling though the SET
is a constant which we will denote $D_0$. However if there is an electron on
dot-1, the rate of tunneling through the SET changes to $D_0+D_1$ with $D_1\
>\ 0$. If $D_0$ does not equal zero then  a current spike (resulting from a
tunnel event in the SET) does not necessarily imply that the electron in the
measured system is in dot-1. In an ideal device the quiescent tunneling
rate, $D_0$ is zero. In reality Johnson noise on the circuit containing the
SET will give a non-zero quiescent tunneling current.

Assuming that the SET island state can be adiabatically eliminated, it is
possible to derive a master equation for the state of the coupled dot
system. This is done in the appendix, and the result is
\begin{equation}  \label{me1}
\frac{d\rho}{dt}=-i[H,\rho]+\gamma_{{\rm dec}}{\cal D}[c_1^\dagger c_1]\rho
= {\cal L}\rho
\end{equation}
where the irreversible part is defined for arbitrary operators $A$ and $B$
by
\begin{equation}
{\cal D}[A]B={\cal J}[A]B - {\cal A}[A]B, \label{uncond}
\end{equation}
where
\begin{eqnarray}
{\cal J}[A]B &=& AB A^\dagger \label{defcalJ}\\
{\cal A}[A]B &=& \frac{1}{2}(A^\dagger AB+B A^\dagger A).
\label{defcalA}
\end{eqnarray}
The decoherence rate is given by
\begin{equation}
\gamma_{{\rm dec}} = 2D_{0}+D_1.
\end{equation}
The fact that the irreversible term is a function of the number operator in
the target qubit is an indication that this describes a QND measurement of
the occupation of the dot-1. It is easy to verify that the stationary
solution of this master equation is an equal mixture of the two accessible
electronic states.

The stochastic record of measurement ideally comprises a sequence of times,
being the times at which electrons tunneled through the SET. In practice of
course these events are not seen due to a finite frequency response of the
circuit (including the SET) which averages each event over some time.
However for the purpose of this paper we will take the zero response time
limit. In this limit the current consists of a sequence of $\delta$ function
spikes. Formally we can write $i(t) = edN/dt$, where $dN(t)$ is a classical
point process which represents the number (either zero or one) of tunneling
events seen in an infinitesimal time $dt$, and $e$ is the electronic charge.
We can think of $dN(t)$ as the increment in the number of electrons $N(t)$
in the collector in time $dt$. It is this variable, the accumulated electron
number transmitted by the SET, which is used in Ref.~\cite{Schoen98}.

The point process $dN(t)$ is formally defined by the conditions
\begin{eqnarray}
[dN(t)]^{2} &=& dN(t)  \label{zo} \\
{\rm E}[dN(t)]/dt &=& D_{0}{\rm Tr}[(1-n_1)\rho_{{\rm c}}(t)(1-n_1)] +
(D_{0}+D_{1}){\rm Tr}[ n_1 \rho_{{\rm c}}(t) n_1 ]  \nonumber \\
&=& D_{0}+ D_{1}\left\langle{\ n_1 }\right\rangle_{{\rm c}}(t) .
\label{mean}
\end{eqnarray}
Here ${\rm E}[x]$ denotes a classical average of a classical stochastic
process $x$, and
\begin{equation}
n_{1} = c_{1}^\dagger c_{1}
\end{equation}
is the occupation number operator for the first dot. The first of these
equations simply expresses the fact that $dN(t)$ equals zero or one. The
second says that the rate of events is equal to a background rate $D_{0}$
plus an additional rate $D_{1}$ if and only if the electron is in the first
dot.

In Eq.~(\ref{mean}), the system state matrix $\rho_{{\rm c}}(t)$ is {\em not}
the solution of the master equation (\ref{me1}). That is because if one has
a record of the current $dN/dt$ through the SET then one knows more about
the system than the master equation indicates. That is to say, $\rho_{{\rm c}%
}(t)$ is actually conditioned by $dN(t^{\prime})$ for $t^{\prime}<t$, hence
the subscript c. The first way of writing Eq.~(\ref{mean}) hints at how $%
dN(t^{\prime})$ conditions $\rho_{{\rm c}}(t)$. From the appendix, the state
at time $t+dt$ given $dN(t)=1$ is
\begin{equation}
\tilde{\rho}_{1}(t+dt) = dt\left[ {\ D_{0}(1-n_1)\rho(t)(1-n_1) +
(D_{0}+D_{1}) n_1 \rho(t) n_1} \right]
\end{equation}
This is an unnormalized state whose norm is equal to the probability of that
event ($dN(t)=1$) occurring, as seen above in Eq.~(\ref{mean}).

The normalized state can be written more elegantly as
\begin{equation}
\rho _{1}(t+dt)=\frac{(D_{0}+D_{1}{\cal J}[n_{1}]+2D_{0}{\cal D}[n_{1}])\rho
(t)}{{\rm Tr}\{(D_{0}+D_{1}{\cal J}[n_{1}]+2D_{0}{\cal D}[n_{1}])\rho (t)\}}=%
\frac{(D_{0}{\cal J}[1-n_{1}]+(D_{0}+D_{1}){\cal J}[n_{1}])\rho (t)}{{\rm E}%
[dN(t)/dt]},  \label{rho1}
\end{equation}
where ${\cal J}$ is as defined above in Eq.~(\ref{defcalJ}).

To write the full conditioned evolution we need to know the density operator
$\rho_{0}(t+dt)$ given that $dN(t)=0$. This can be found from Eq.~(\ref{rho1}%
) plus the fact that when averaged over the observed classical point process
$dN$,
\begin{equation}
\tilde\rho_{0}(t+dt) + \tilde\rho_{1}(t+dt) = (1+{\cal L}dt) \rho(t).
\end{equation}
That is to say, {\em on average} the system still obeys the master equation (%
\ref{me1}). From this equation, we obtain
\begin{eqnarray}  \label{rho0}
\tilde\rho_{0}(t+dt) &=& \rho(t)-dt\{D_{0}{\cal A}[1-n_1]\rho(t) + (D_{0}+
D_{1}){\cal A}[n_1]\rho(t)+i[H,\rho(t)]\}, \\
&=& \rho(t)-dt\left\{ {D_{0}\rho(t)+D_{1}{\cal A}[n_1]\rho(t)-i[H,\rho(t)]} \right\},
\end{eqnarray}
where ${\cal A}$ is as defined above in Eq.~(\ref{defcalA}).
 Once again this state is
unnormalized and its norm gives the probability that $dN(t)=0$, that is
\begin{equation}
{\rm Tr}[\tilde\rho_{0}(t+dt)] = 1 - {\rm E}[dN(t)]
\end{equation}

Using the variable $dN(t)$ explicitly, the conditioned state at time $t+dt$
is
\begin{equation}
\rho_{c}(t+dt) = dN(t) \frac{\tilde{\rho}_{1}(t+dt)}{{\rm Tr}[\tilde{\rho}%
_{1}(t+dt)]} + [1-dN(t)]\frac{\tilde{\rho}_{0}(t+dt)}{{\rm Tr}[\tilde{\rho}%
_{0}(t+dt)]}.
\end{equation}
Since $dN(t)$ is almost always zero we can set $dN(t)dt=0$ and expand this
expression to finally obtain the stochastic master equation, conditioned on
the observed event in time $dt$
\begin{eqnarray}
d\rho_{{\rm c}} & = & dN(t)\left [\frac{D_0+D_1{\cal J}[n_1]
+2D_{0}{\cal D}[n_1]}{D_0+D_1 {\rm Tr}[\rho_{{\rm c}} n_1]}%
-1\right ]\rho_{{\rm c}} \nonumber \\
& & \mbox{}+\, dt\left\{-D_1{\cal A}[n_1]\rho_{{\rm c}}+D_1{\rm %
Tr}[\rho_{{\rm c}} n_1]\rho -i[H,\rho_{{\rm c}}]\right\}
\label{condev}
\end{eqnarray}
Note that averaging this equation over the observed stochastic process (by
setting $dN$ equal to its  expected value) gives the unconditional master
equation (\ref{me1}).

\section{Average steady state properties}

We now analyze in some detail the ensemble averaged properties of the system
based on the unconditional master equation. In particular we calculate the
stationary noise power spectrum of the current through the SET when there is
the possibility of coherent tunneling between dot-2 and the measured dot-1.
The details of how the quantum stochastic processes in the SET determine the
average current though the SET are given in reference \cite{Sun99}. The link
with the stochastic formalism of the preceding section is that the current $%
i(t)$ through the SET is given by
\begin{equation}
i(t) = e\frac{dN(t)}{dt}.
\end{equation}

First we calculate the steady state current
\begin{eqnarray}
i_{\infty} &=& {\rm E}[i(t)]_\infty  \nonumber \\
& = & e(D_0+D_1\langle c_1^\dagger c_1\rangle_{\infty}) \\
& = & e(D_0+\frac{D_1}{2}),  \label{current}
\end{eqnarray}
where the $\infty$ subscript indicates that the system is at steady-state.
The fluctuations in the observed current, $i(t)$ are quantified by the
two-time correlation function:
\begin{eqnarray}
G(\tau )&=& {\rm E}[i(t)i(t+\tau) - i_\infty^2]_\infty \\
&=& e i_{\infty }\delta (\tau )+\left\langle i(t),i(t+\tau )\right\rangle
_{\infty }^{\tau \neq 0} . \label{Eq-corr}
\end{eqnarray}
Here $\left\langle{A,B}\right\rangle \equiv \left\langle{AB}%
\right\rangle-\left\langle{A}\right\rangle\left\langle{B}\right\rangle$. The
fact that the multiplier of the shot noise is $ei_{\infty}$ rather than the
usual $(e/2)i_{\infty}$ is because of the approximation we have made in
treating the SET. Specifically, we have adiabatically eliminated the SET by
taking the limit where any electron which tunnels onto the SET island from
the emitter  immediately tunnels off to the collector. This means that, on
the  time scales we are interested in, there is a perfect correlation
between the emitter current and collector current. This leads to a  doubling
of the shot noise level. Of course, at very high  frequencies, higher than
we are interested in, the true shot noise  level of $(e/2)i_{\infty}$ could
still be seen in principle.

To relate these classical averages to the fundamental quantum processes
occurring in the well we apply the theory of open quantum system \cite
{Carmichael93,Wiseman93} to the present system. Specifically, we can relate
the correlation function for the current to the following quantum averages
\begin{equation}
\left\langle {i(t)i(t+\tau )}\right\rangle _{\infty }^{\tau \neq 0}=e^{2}%
{\rm Tr}\left[ {(D_{0}+D_{1}{\cal J}[n_{1}]+2D_{0}{\cal D}[n_{1}])e^{{\cal L}%
\tau}(D_{0}+D_{1}{\cal J}[n_{1}]+2D_{0}{\cal D}[n_{1}])\rho _{\infty
}}\right].
\end{equation}
Because $\rho _{\infty }$ is an equal mixture of the two electron states, it
satisfies
\begin{equation}
{\cal D}[n_{1}]\rho _{\infty }=0.
\end{equation}
In addition, the following identies for arbitrary operators $A$ and $B$ are
easy to prove: ${\rm Tr}\left[ {{\cal D}[A]B}\right] \equiv 0$, ${\rm Tr}%
\left[ {{\cal J}[n_{1}]B}\right] \equiv {\rm Tr}[n_{1}B]$, ${\rm Tr}[e^{%
{\cal L}\tau}B]={\rm Tr}[B]$, and ${\rm Tr}[Ae^{{\cal L\tau }}\rho _{\infty
}]=%
{\rm Tr}[A\rho _{\infty }]$. Using these simplfications we obtain
\begin{equation}
\left\langle {i(t),i(t+\tau )}\right\rangle _{\infty }^{\tau \neq
0}=D_{1}^{2}e^{2}\left\{ {{\rm Tr}\left[ {n_{1}e^{{\cal L\tau }}{\cal J}%
[n_{1}]\rho _{\infty }}\right] -{\rm Tr}[n_{1}\rho _{\infty }]^{2}}\right\} .
\end{equation}

Evaluating this expression we find
\begin{equation}
G(\tau )=ei_{\infty }\delta ({\tau })+\frac{e^{2}D_{1}^{2}}{8}\left( \frac{%
\mu_{+}e^{\mu _{-}\tau }-\mu_{-}e^{\mu_{+}\tau }}{\sqrt{%
(\gamma _{{\rm dec}}/4)^{2}-\Omega ^{2}}}\right) ,
\end{equation}
where
\begin{equation}
\mu_{\pm }=-(\gamma _{{\rm dec}}/4)\pm \sqrt{(\gamma _{{\rm dec}%
}/4)^{2}-\Omega ^{2}},
\end{equation}
and where the first term represents the shot noise component as discussed
above. The power spectrum of the noise is
\begin{equation}
S(\omega )=\int_{0}^{\infty }d\tau G(\tau )2\cos (\omega \tau ),
\end{equation}
which evaluates to
\begin{equation}
S(\omega )=ei_{\infty }+\frac{e^{2}D_{1}^{2}\Omega ^{2}/2}{\sqrt{(\gamma _{%
{\rm dec}}/4)^{2}-\Omega ^{2}}}\left\{ \frac{1}{\mu_{+}^{2}+\omega ^{2}}%
-\frac{1}{\mu_{-}^{2}+\omega ^{2}}\right\} .
\end{equation}

In the case that $\Omega > \gamma_{{\rm dec}}/4$ the spectrum will have a
double peak structure indicating that coherent tunneling is taking place
between the two coupled dots. For smaller $\Omega$ only a single peak
appears in the spectrum. We can thus use the noise power spectrum of the
current though the SET as a means to measure the tunnel coupling between
dots if the tunnel coupling is high enough. We illustrate this in figure \ref
{fig3}.

\section{Analytical Results for Conditional Dynamics}

We now return to the stochastic master equation for the conditioned state,
\begin{eqnarray}
d\rho_{{\rm c}} & = & dN\left [\frac{D_0+D_1{\cal J}[c_1^\dagger c_1]+2D_{0}%
{\cal D}[c_{1}^\dagger c_{1}]}{D_0+D_1 \mbox{Tr}[\rho_{{\rm c}} c_1^\dagger
c_1]}-1\right ]\rho_{{\rm c}} \nonumber \\
& & \mbox{}+dt\left\{-D_1\frac{1}{2}\{c_1^\dagger c_1,\rho_{{\rm c}}\}+D_1%
\mbox{Tr}[\rho_{{\rm c}} c_1^\dagger c_1]\rho -i[H,\rho_{{\rm
c}}]\right\}.
\label{condev2}
\end{eqnarray}
Comparing this to the unconditional master equation
\begin{equation}
\dot{\rho} = -i[H,\rho]+\gamma_{{\rm dec}}{\cal D}[c_{1}^\dagger c_{1}]\rho
\end{equation}
we see that decoherence between the two coupled dots, 1 and 2, takes place
at the rate $\gamma_{{\rm dec}}=2D_{0}+D_{1}$, but that the system decides
between the two possibilities (electron on dot-1 or on dot-2) on a time
scale that depends on $D_{1}$ and $D_0$ in some more complicated way. Of
course this measurement time scale is necessarily at least as large as the
decoherence time scale because successfully distinguishing between the two
dots would by definition destroy any coherence between them.

The different measurement time scales can be derived most easily by
introducing the Bloch representation of the state matrix:
\begin{equation}  \label{bloch}
\rho=\frac{1}{2}\left (I+x\sigma_x+y\sigma_y+z\sigma_z\right )
\end{equation}
where the Pauli matrices are defined using the Fermi operators for the two
dots
\begin{eqnarray}
\sigma_x & = & c^\dagger_{1}c_{2}+c^\dagger_{2}c_{1} \\
\sigma_y & = & -ic^\dagger_{1}c_{2}+ic^\dagger_{2}c_{1} \\
\sigma_z & = & c^\dagger_{2}c_{2}-c^\dagger_{1}c_{1}
\end{eqnarray}
In this representation the means of the Pauli matrices $\sigma_\alpha$ are
given by the respective coefficient $\alpha$, with $\alpha=x,y,z$.

The stochastic master equation can now be written as a set of coupled
stochastic differential equations for the Bloch sphere variables as
\begin{eqnarray}
dz_{{\rm c}} &=&\Omega x_{{\rm c}}dt+\frac{D_{1}}{2}(1-z_{{\rm c}}^{2})dt-%
dN(t)\frac{D_{1}(1-z_{c}^{2})/2}{D_{0}+D_{1}(1-z_{c})/2}
\\
dx_{{\rm c}} &=&-\Omega z_{{\rm c}}dt-\frac{D_{1}}{2}z_{{\rm c}}x_{{\rm c}%
}dt-dN(t)x_{\rm c} \\
dy_{{\rm c}} &=&-\frac{D_{1}}{2}z_{{\rm c}}y_{{\rm c}}dt-dN(t)y_{\rm c}.
\end{eqnarray}
Again the c-subscript is to emphasize that these variables refer to the
conditional state. If we average over the noise, the ensemble dynamics is
then seen to be given by
\begin{eqnarray}
\frac{dz}{dt} &=&\Omega x \\
\frac{dx}{dt} &=&-\Omega z-\frac{\gamma _{{\rm dec}}}{2}x \\
\frac{dy}{dt} &=&-\frac{\gamma _{{\rm dec}}}{2}y
\end{eqnarray}
where $\alpha ={\rm E}[\alpha _{{\rm c}}]$ denotes the averaging over the
ensemble
of conditional states. These equations are exactly what would be obtained
directly from the ensemble averaged master equation Eq(\ref{me1}). In
particular we note that the average population difference $z$ between the
dots is a constant of the motion in the absence of any free Hamiltonian.
However the stochastic differential equations enable us to calculate
important averages that are {\em not} obtainable from the master equation.
For example, if the model does indeed describe a measurement of $%
c_{1}^{\dagger }c_{1}=(1-\sigma _{z})/2$, then, in the absence of tunneling,
we would expect to see the conditional state become localized at either $z=1$
or $z=-1$. Indeed for $\Omega =0$ we can see from the conditional equation
for $z_{{\rm c}}$ that $z_{{\rm c}}=\pm 1$ is a fixed point.

We can take into account both fixed points by considering $z_{{\rm c}}^{2}$.
In the absence of tunneling this must must approach $1$ for all
trajectories, since the system will eventually become localized due to the
measurement in one dot or the other. Therefore it is sensible to take the
ensemble average ${\rm E}[z_{{\rm c}}^{2}]$ and find the rate at which this
deterministic quantity approaches one. Noting that for a stochastic variable
we have $d(z^{2})=2zdz+dzdz$, and that ${\rm E}[dN^{2}]={\rm E}%
[dN]=[D_{0}+D_{1}(1-z_{{\rm c}})/2]dt$, we find that
\begin{equation}
\frac{d{\rm E}[z_{{\rm c}}^{2}]}{dt}={\rm E}\left[ \frac{D_{1}^{2}(1-z_{{\rm %
c}}^{2})^{2}}{4D_{0}+2D_{1}(1-z_{{\rm c}})}\right] .
\end{equation}

If the system starts state which has an equal probability for single
electron to be on each dot then $z_{{\rm c}}(0)=0$ and in the ensemble
average this would remain the case. However if we ensemble average $z_{{\rm c%
}}^2$ over many quantum trajectories then for short times we find
\begin{equation}
{\rm E}[z_{{\rm c}}^2(\delta t)]=\frac{D_1^2}{4D_0+2D_1}\delta t  \label{z2t}
\end{equation}
That is to say, the system tends towards a definite state (with $z_{{\rm c}%
}=\pm 1$ so $z_{{\rm c}}^{2}=1$) at an initial rate of $%
D_{1}^{2}/(4D_{0}+2D_{1})$. For vanishing $D_{0}$, this is the same as the
decoherence rate, $D_{1}/2$, as expected. But for $D_{0} \gg D_{1}$, the
rate goes to $(D_{1}/2D_{0})\times D_{1}/2 \ll D_{1}/2$. That is, the rate
at which the system becomes localized at one or the other dot is much less
than the decoherence rate. This result cannot be obtained from the ensemble
averaged master equation alone. It is a direct reflection of the fact that
for $D_0\neq 0$ a tunneling event cannot be unambiguously attributed to the
location of an electron on the double dot system. As the rate of
localization is a direct indication of the quality of the measurement, we
can use the localization rate defined as
\begin{equation}
\gamma_{{\rm loc}}=\frac{D_1^2}{4D_0+2D_1}
\end{equation}
as an important parameter defining the quality of the measurement. This
parameter is related to the signal-to-noise ratio for this measurement as we
now show.

For a Poisson process at rate $R$, the probability for $m$ events to occur
in time $T$ is
\begin{equation}
p(m;T)=\frac{(R T)^m}{m!}e^{-R T}
\end{equation}
The mean and variance of this distribution are equal and given by ${\rm E}%
(m) = {\rm Var}(m) = R T$. Now consider an electron which is, with equal
likelihood, in either dot, so that $z=0$. If the electron is in dot 1 then
the rate of electrons passing through the SET is $D_{0}+D_{1}$; if it is in
dot two then it is just $D_{0}$. These two possibilities will begin to be
distinguishable when the difference in the means of the two distributions $%
p(m,T)$ is of order the square root of the sum of the variances. That is,
when
\begin{equation}
D_{1}T \sim \sqrt{D_{0}T + (D_{0} + D_{1})T}
\end{equation}
Solving this for $T$ gives a characteristic rate
\begin{equation}
T^{-1} \sim \frac{D_{1}^{2}}{2D_{0}+D_{1}}.
\end{equation}
The right hand side of this expression is simply twice the $\gamma_{{\rm loc}%
}$ defined above. A similar conclusion
is reached in reference \cite{Schoen98}.

In the ideal limit of no quiescent current in the SET $D_0=0$, the
stochastic master equation can be replaced by a stochastic Schr\"{o}dinger
equation, and will collapse to a single possibility at a rate $D_1$ which is
the same as the decoherence rate. The effect of $D_{0}$ is most clearly seen
in the other limit, $D_1 \ll D_0$, as noted above. In this limit the single
electron makes only a small relative change in the tunneling rate through
the SET. As the rate of jumps also becomes large, the trajectories in
this limit take on the appearance of diffusion rather than jumps.
The rate
at which the electron localizes into one well or the other scales as $%
D_1^{-2}D_0$, which is
much longer than the decoherence time scale $D_0^{-1}$.
\label{earlier}

\section{Numerical Simulations of Conditional dynamics}

\label{later}

We now turn to numerical simulations of the conditional evolution and to
estimate the conditions for a good measurement. Unlike traditional condensed
matter measurements we wish to describe repeated measurements made on a
single quantum system rather than a single measurement made upon an ensemble
of systems. To do this we use  the conditional dynamics of the measured
system given a particular measurement record as described by the above
stochastic evolution equation (\ref{condev2})

We return to the Bloch description defined in Eq.~(\ref{bloch}). In what
follows we will assume that $y(0)=0$. For the form of tunneling used here
the value of $y$ does not in fact change under either conditional or
ensemble averaged dynamics. If the conditional state of the system remains
in a pure state then $x_{{\rm c}}^2+y_{{\rm c}}^2+z_{{\rm c}}^2=x_{{\rm c}%
}^2+z_{{\rm c}}^2=1$. As noted previously this can only occur if the bare
tunneling rate ($D_0$) is zero, when a tunneling event can unambiguously be
attributed to the occupation of the target dot and no information is lost
about the state of the system. In the more realistic case in which $D_0\neq 0
$, we can use the quantity $x_{{\rm c}}^2+z_{{\rm c}}^2$ as a measure of the
purity of the sate, or equivalently as a measure of how much information the
conditional record of measurements gives about the actual state of the two
coupled dots. If the conditional state is a maximally mixed state of a two
state system then $x_{{\rm c}}^{2}+z_{{\rm c}}^{2}=0$.  We now describe in
detail the numerical simulation of the conditional dynamics.

\subsection{No Background Current}

First we consider the case $D_{0}=0$, so the system is always in a pure
state. Typical trajectories are shown in Fig.~4 for various values of $\Omega
$. For small $\Omega \ll D_{1}/2$ we see little evidence for coherent
tunneling. Most of the time the electron is localized almost entirely in one
well or the other. However, there is an asymmetry between the wells. A
transition from dot-2 into dot-1 (the target) is sudden, occurring whenever
an electron tunnels through the SET. A transition the other way takes a time
of order $2/D_{1}$. This time is still much smaller than the average time
between state-changing transitions, which can be shown analytically to be $%
D_{1}/\Omega^{2}$. Thus over a long time, as shown in Fig.~4(a), the system
still has the appearance of a random telegraph. This behaviour gives rise to
a single-peaked noise spectrum, as shown in Fig.~3(a).

For moderate $\Omega \sim D_{1}/2$ the system is no longer well-localized in
one dot or the other. Rather, the dynamics is complicated with clearly
non-sinusoidal oscillations from one dot to the other interspersed with
jumps into the target dot. The fact that oscillations are present gives
peaks in the current noise spectrum, as shown in Fig.~3(b). The position of
these peaks is a frequency less than $\Omega$, as shown analytically in
Sec.~III.

For $\Omega \gg D_{1}/2$ the dynamics once again becomes simple, with nearly
sinusoidal oscillations interspersed with jumps which occur with an average
rate of $D_{1}/2$. This corresponds to a noise spectrum having a very sharp
feature at $\omega \approx \pm \Omega$, as shown in Fig.~3(c).

The change in behaviour as $\Omega$ increases is summarized in Fig.~5. There
we plot E$[z_{{\rm c}}^{2}]$ versus $\Omega/D_{1}$. The quantity E$[z_{{\rm c%
}}^{2}]$ measures how well localized the electron is at one well or the
other, and would be $1$ if the electron were always localized and $0$ if it
were never localized. It is actually possible to calculate this quantity
numerically without using a stochastic ensemble, as follows.

With no background current, every time an electron tunnels through the SET
the electron on the dots is known to be on dot $1$. If there are no further
SET tunneling events for a time $t$ later then from Eq.~(\ref{rho0}), the
system evolves up to that time by the equation
\begin{equation}
d\tilde\rho_{0}(t) = -dt\{D_{1}\{c_{1}^\dagger c_{1},\tilde\rho%
_{0}(t)\}/2+i[H,\tilde\rho_{0}(t)]\}.
\end{equation}
Because there is no background current, and because the initial state is
pure, it is possible to rewrite this in terms of a non-Hermitian
Schr\"odinger equation
\begin{equation}  \label{sse}
d|{\tilde\psi_{0}(t)}\rangle = -dt(iH + D_{1}c_{1}^\dagger c_{1}/2)|{\tilde%
\psi_{0}(t)}\rangle.
\end{equation}
Here it must be remembered that the norm of this state represents the
probability for the event that no electron has passed through the SET 
since the last one a time $t$ ago:
\begin{equation}
p_{0}(t) = {\rm Tr}[\tilde\rho_{0}(t)] = \langle{\tilde\psi_{0}(t)}|{\tilde%
\psi_{0}(t)}\rangle.
\end{equation}

It is not difficult to show that the solution to Eq.~(\ref{sse}) satisfying
the initial condition $|{\psi_{0}(0)}\rangle=|{1,0}\rangle$ is
\begin{equation}
|{\tilde\psi_{0}(t)}\rangle = \alpha(t)|{1,0}\rangle + \beta(t)|{0,1}\rangle,
\end{equation}
where the occupation numbers refer to the dots one and two in order. Here $%
\alpha$ and $\beta$ are real numbers defined by
\begin{eqnarray}
\alpha(t) &=& \frac{1}{\lambda_{+}-\lambda_{-}}\left( {\
\lambda_{+}e^{\lambda_{+}t} - \lambda_{-}e^{\lambda_{-}t}} \right) , \\
\beta(t) &=& \frac{-\Omega/2}{\lambda_{+}-\lambda_{-}}\left( {\
e^{\lambda_{+}t} - e^{\lambda_{-}t}} \right) ,
\end{eqnarray}
where
\begin{equation}
\lambda_{\pm} = \frac{1}{2} \left[ {\ -\frac{D_{1}}{2} \pm \sqrt{\left( {%
\frac{D_{1}}{2}} \right)^{2} - \Omega^{2}}} \right]
\end{equation}
The conditioned quantum expectation value for $\sigma_{z}$ is
\begin{equation}
z_{0}(t) = \frac{\langle{\tilde\psi_{0}(t)}|\sigma_{z}|{\tilde\psi_{0}(t)}%
\rangle} {p_{0}(t)} = \frac{\beta^{2}-\alpha^{2}}{\beta^{2}+\alpha^{2}}
\end{equation}

Now in steady state the probability $p_{0}(t)$ that there is no increment in
$N(t)$ for a time $t$ ago is related to the probability $q_{0}(t)$ that the
last increment was a time $t$ ago by
\begin{equation}
q_{0}(t) = \frac{p_{0}(t)}{\int_{0}^{\infty}p_{0}(s)ds}.
\end{equation}
Since at steady state all conditioned states are uniquely identified by how
long it has been since the last SET event, the ensemble average for $%
z_{c}^{2}$ is simply given by
\begin{eqnarray}
{\rm E}[z_{c}^{2}] &=& \int_{0}^{\infty} q_{0}(t) [z_{0}(t)]^{2}dt \\
&=&\left[ {\int_{0}^{\infty} \left( {\beta^{2}+\alpha^{2}} \right)dt}
\right]^{-1} {\int_{0}^{\infty}\frac{\left( {\beta^{2}-\alpha^{2}}
\right)^{2}} {\beta^{2}+\alpha^{2}} dt } .
\end{eqnarray}
Unfortunately it does not appear possible to evaluate the second integral
here analytically. However a numerical integration is easy. The results, shown in Fig.~5, 
is in  agreement with the ensemble
averages obtained numerically using the stochastic
master equation. 

\subsection{A Finite Background Current}

We next consider the case where $D_{0}\neq 0$. We show two plots, both with $%
\Omega=D_{1}$, which is a regime in which coherent tunneling is clearly
evident in the current noise spectrum. The first plot, in Fig.~6, is for $%
D_{0}=D_{1}$. Here coherent oscillations are still evident in $z$, but $z$
rarely attains its extreme values of $\pm 1$. The conditioned state is no
longer pure, even immediately after a count. Also, the conditioned state
following a count now depends on the state before the count. For this reason
an exact solution by the method of the preceding section is impossible.

The second plot, in Fig.~7, is for $D_{0}=10 D_{1}$. Here coherent
oscillations are no longer obvious in the condition mean of $\sigma_{z}$,
even thought they are present in the spectrum (as small features above the
shot noise), as calculated in Sec.~III. In this regime $D_{0}\gg D_{1}$ so
the diffusive limit discussed at the end of Sec.~\ref{earlier}
applies.
%Moreover,
%with this choice of $\Omega$, the condition $\Omega^{2} \gg D_{1}^{3}/D_{0}$
%is also satisfied, so the linearized results of Eq.~(\ref{linres}) also
%apply. This is because the state of the electron is almost completely mixed,
%with the current being too noisy to collapse the electron into a pure
%quantum state.

\section{Discussion and Conclusion}

The three parameters we need to compare our theoretical results with
experiment are $\Omega$, $D_0$ and $D_1$. The two incoherent tunneling rates
can be obtained by considering how they determine the steady state current
through the device. This is given in Eq~(\ref{current}). Our model
implicitly assumes that the quiescent noise in the SET is shot-noise
limited, based as it is on elementary tunneling events. However from the
point of view of the macroscopic circuit in which the SET is placed, the
tunnel junctions appear as a capacitor in series with a resistor ( see note
14 in reference \cite{Schoelkopf98}). If the resistance of the junction is $R
$ and the capacitance is $C$ the fundamental time constant for the junction
is $\frac{1}{RC}$. This sets an upper bound for the tunneling rate $%
\gamma\leq \frac{1}{RC}$ For a typical Al/AlO$_{x}$/Al junctions we have $C\approx 0.2f$F and $R\approx 50 k\Omega$ and
thus the time constant is $\gamma\approx 10^{11}\ \mbox{s}^{-1}$. For a
double tunnel junction device, the maximum conductance is achieved for a
symmetric pair of tunnel junctions. The value of $D_1$ in this case is given
by $D_1=\gamma/2$ where $\gamma$ is the tunneling rate of the SET under
conditions of maximum conductance (see appendix). We thus estimate that $%
D_1\approx 5 \times 10^{10}\ \mbox{s}^{-1}$. The background tunneling rate
through the SET depends on temperature as well as the bias conditions (see
appendix). Typical maximum and minimum conductance for the SET at different
temperatures have been measured by Joyez {\em et al.} \cite{Joyez97}. At a
temperature of $100 m\mbox{K}$ the minimum conductance is approximately
zero, and thus at this temperature we can safely take $D_0=0$, the zero
temperature result. However at a temperature of $400 \mbox{mK}$ the ratio of the maximum to
minimum conductance is 2.2.  This indicates that at 400 mK, $D_0=4 \times 10^{10} \mbox{s}^{-1}$.

The value we choose for the tunneling rate depends strongly on the
particular quantum dot system. We will consider the value appropriate for
the single electron measurement scheme of Kane {\em et al.}\cite{Kane99}. In
this model the two localized states correspond to an electron on a tellurium
ion donor or at a nearby interface below the donor. The tunneling rate is
then between the donor state and the interface state. Kane {\em et al.}
estimate that for this system $\Omega\approx 10^9\mbox{s}^{-1}$. This value
is too low to observe a peak in the noise power spectrum away from zero
frequency. However the tunneling rate could be increased by changing the
donor interface bias voltage.

In conclusion we have presented a simple model to describe single electron
measurements on a coupled double quantum dot system using an SET. We have
given both the stationary (ensemble averaged) properties of the current
through the SET as well as the conditional dynamics of repeated measurements
on a single system. This illustrates how quantum trajectory methods may be
naturally adapted to single electronics, and aid in the interpretation of
ensemble averaged properties. We believe these models will prove useful in
current attempts to fabricate quantum logic gates in solid state devices.

\appendix

\section{Derivation of the master equation.}

It will suffice to consider a single quantum dot near the SET. This allows
us to remove any reference to the electron field labeled by $c_2,\
c_2^\dagger$. The SET is modeled as a single biased double barrier (single
well) device with a single bound state on the well described by the Fermi
operators $b,\ b^\dagger$. The total Hamiltonian for the system including
the reservoirs is
\begin{equation}
H=H_0+H_{CB}+H_{RT}+H_{LT}
\end{equation}
The term $H_{CB}$ is the Coulomb blockade term and is given by\cite{Schoen97}
\begin{equation}
H_{CB}=\hbar\chi c_1^\dagger c_1 b^\dagger b
\end{equation}
where $\hbar\chi$ is the Coulomb blockade energy gap (see figure \ref{fig2}%
). Note this term can only be nonzero if there is an electron on the island
and on the dot, in which case the energy of the island electron is
increased. The terms $H_{RT},\ H_{LT}$ described the tunneling between the
many modes in the left and right ohmic contacts and the bound state on the
SET \cite{Sun99}
\begin{eqnarray*}
H_{LT} & = & \sum_k T_{Lk}a_{Lk}^\dagger b+T_{Lk}^*a_{Lk} b^\dagger \\
H_{RT} & = & \sum_k T_{Rk}a_{Rk}^\dagger b+T_{Rk}^*a_{Rk} b^\dagger
\end{eqnarray*}
where $a_{Lk},\ a_{Rk}$ are respectively the Fermi field annihilation
operators for the left and right reservoir states at momentum $k$. The
tunneling matrix elements between respectively the left and right Ohmic
contacts and the island are $T_{Lk},\ T_{Rk}$. The free Hamiltonian for the
the system is
\begin{equation}
H_0=\hbar\sum_k \omega_k^L a_{Lk}^\dagger a_{Lk} +\omega_k^R a_{Rk}^\dagger
a_{Rk}+\hbar\omega_1c_1^\dagger c_1+\hbar\omega_0 b^\dagger b
\end{equation}
We now transform to an interaction picture to remove the terms $H_0+H_{CB}$.
The dynamics in the Schr\"{o}dinger picture is now described by the time
dependent Hamiltonian
\begin{eqnarray*}
H_I(t) & = & \sum_k T_{Lk}a_{Lk}^\dagger be^{i\chi t c_1^\dagger
c_1}e^{-i(\omega_k^L-\omega_0)t} \\
& & +T_{Lk}^*a_{Lk} b^\dagger e^{-i\chi tc_1^\dagger
c_1}e^{i(\omega_k^L-\omega_0)t} \\
& & T_{Rk}a_{Rk}^\dagger be^{i\chi tc_1^\dagger
c_1}e^{-i(\omega_k^R-\omega_0)t} \\
& & +T_{Rk}^*a_{Rk} b^\dagger e^{-i\chi tc_1^\dagger
c_1}e^{i(\omega_k^R-\omega_0)t}
\end{eqnarray*}
Using the fact that $(c_1^\dagger c_1)^n=c_1^\dagger c_1$ we find
\begin{eqnarray*}
H_I(t)=H_1(t)+H_2(t)
\end{eqnarray*}
where
\begin{eqnarray*}
H_1(t) & = & (1-c_1^\dagger c_1)\sum_k (T_{Lk}a_{Lk}^\dagger
be^{-i(\omega_k^L-\omega_0)t}+{\rm H.c.})+(T_{Rk}a_{Rk}^\dagger
be^{-i(\omega_k^R-\omega_0)t}+{\rm H.c.}) \\
H_2(t) & = & c_1^\dagger c_1\sum_k (T_{Lk}a_{Lk}^\dagger
be^{-i(\omega_k^L-\omega_0-\chi)t}+{\rm H.c.})+(T_{Rk}a_{Rk}^\dagger
be^{-i(\omega_k^R-\omega_0-\chi)t}+{\rm H.c.})
\end{eqnarray*}
Notice that if there is no electron on the dot and $c_1^\dagger
c_1\rightarrow 0$ then the second term is zero and the first term is a
standard tunneling interaction onto a bound state with energy $\hbar\omega_0$%
. On the other hand if there is an electron on the dot $c_1^\dagger
c_1\rightarrow 0$ and the first term is zero and the second term is a
standard tunneling interaction onto a bound state with energy $%
\hbar(\omega_0+\chi)$ as expected.

The derivation of the master equation for the state matrix $R$ for the
system (SET and quantum dot) can now proceed using standard techniques which
we will sketch. The objective is to obtain a semigroup evolution in Lindblad
form (that is to say positivity-preserving irreversible dynamics) for the
state of the SET island and the dot alone with no reference to the ohmic
contacts. The ohmic contacts are treated as perfect Fermi thermal reservoirs
with a very fast relaxation constants. Each ohmic contact (left and right)
remains in thermal equilibrium with chemical potentials $\mu_L,\ \mu_R$, but
the total system is not in thermal equilibrium due to the external bias
potential, $V$ with $eV=\mu_L-\mu_R$ (see references \cite{datta,milburn}
for further discussion). We first define a time interval $\delta t$ which is
slow compared to the dynamics of the island and the dot but very long
compared to the time scale in which the ohmic contacts relax back to their
steady state. The change in the state matrix $W$ of the system (SET and dot)
and environment (Ohmic contacts) from time $t$ to $t+\delta t$, to second
order in the tunnel coupling energy, is given by
\begin{equation}
W(t+\delta t)=W(t)-i\delta t[H_I(t),W(t)]-\delta t\int_t^{t+\delta
t}dt_1[H_I(t),[H_I(t_1),W(t_1)]]
\end{equation}
We now make the first Markov approximation and assume that at any time the
state of the total system may be approximated by $W(t)=R(t)\otimes\rho_L%
\otimes\rho_R$ , that is to say the left and right ohmic contacts
instantaneously relax back to Fermi distributions. We now obtain an
evolution equation for $R(t)$, the state of the island and the dot by
tracing over the reservoirs. The result is
\begin{eqnarray*}
\frac{dR(t)}{dt} & = & \left [\gamma_L(1-f_L(\omega_0))+\gamma_R(1-f_R(%
\omega_0)\right ]{\cal D}[b(1-c_1^\dagger c_1)]R \\
& & +\left [\gamma_L f_L(\omega_0)+\gamma_Rf_R(\omega_0)\right ]{\cal D}%
[b^\dagger(1-c_1^\dagger c_1)]R \\
& & +\left [\gamma^{\prime}_L(1-f_L(\omega_0+\chi))+\gamma^{\prime}_R(1-f_R(%
\omega_0+\chi))\right ]{\cal D}[bc_1^\dagger c_1]R \\
& & +\left [\gamma^{\prime}_Lf_L(\omega_0+\chi)+\gamma^{\prime}_Rf_R(%
\omega_0+\chi)\right ]{\cal D}[b^\dagger c_1^\dagger c_1]R
\end{eqnarray*}
where for arbitrary operators $A$ and $B$, ${\cal D}[A]B=AB A^\dagger -\frac{%
1}{2}(A^\dagger A B-BA^\dagger A)$ and where $f_{L,R}(\omega)$ is the Fermi
filling probability for the left/right ohmic contact at the energy $%
\hbar\omega$. The rates $\gamma_{L,R}$ and $\gamma^{\prime}_{L,R}$ determine
the rate of injection from the left ohmic contact into the island or
emission from the island into the right ohmic contact under the conditions
of no electron on the dot (unprimed) and with an electron on the dot
(primed). These are evaluated using the second markov approximation as
\begin{eqnarray}
\gamma_L &=& |T_{Lk_0}|^2, \\
\gamma_R &=&|T_{Rk_0}|^2, \\
\gamma^{\prime}_{L} &=& |T_{Lk_0^{\prime}}|^2 \\
\gamma^{\prime}_{R} &=& |T_{Rk_0^{\prime}}|^2
\end{eqnarray}
where $k_0=\sqrt{2m\omega_0/\hbar}$ and $k_{0}^{\prime}=\sqrt{%
2m(\omega_0+\chi)/\hbar}$.

The ideal quiescent state of the SET is defined as $f_L(\omega_0)=1,\
f_R(\omega_0)=1$ while $f_L(\omega_0+\chi)=1,\ f_R(\omega_0+\chi)=0$. Under
these conditions the master equation reduces to
\begin{equation}
\frac{dR}{dt} = \gamma_{R}{\cal D}[b(1-c_1^\dagger c_1)]R+\gamma_L{\cal D}%
[b^\dagger (1-c_1^\dagger c_1)]R+\gamma^{\prime}_R{\cal D}[b c_1^\dagger
c_1]R + \gamma^{\prime}_L{\cal D}[b^\dagger c_1^\dagger c_1]R
\end{equation}

We now wish to derive a master equation for the state matrix $\rho$ for the
dot alone. This is easiest if we assume that $\gamma_{R},\gamma_{R}^{\prime}$
are much larger than all other rates in the system. In this case it is
possible to adiabatically eliminate the SET island using techniques similar
to that in Ref.~\cite{Wis93a}. We expand the state matrix $R$ in powers of $%
1/\gamma_{R}$ or $1/\gamma_{R}^{\prime}$ as
\begin{equation}
R = \rho_{0}\otimes|{0}\rangle\langle{0}| + \rho_{1}\otimes|{0}\rangle\langle%
{0}|.
\end{equation}
The equations of motion for $\rho_{1}$ and $\rho_{0}$ are
\begin{eqnarray}
	\dot{\rho}_{1} & = & -\gamma_{R}{\cal A}[1-n_{1}]\rho_{1} +
	\gamma_{L}{\cal J}[1-n_{1}]\rho_{0}-\gamma'_{R}{\cal
A}[n_{1}]\rho_{1} +
	\gamma'_{L}{\cal J}[n_{1}]\rho_{0}
	\label{dotrho1}  \\
	\dot{\rho}_{0} & = & \gamma_{R}{\cal J}[1-n_{1}]\rho_{1} -
	\gamma_{L}{\cal A}[1-n_{1}]\rho_{0} + \gamma'_{R}{\cal J}[n_{1}]\rho_{1} -
	\gamma'_{L}{\cal A}[n_{1}]\rho_{1}
	\label{dotrho0}
\end{eqnarray}
Here $n_{1}=c_{1}^{\dagger} c_{1}$ and ${\cal J}$ and ${\cal A}$ are as
defined in Eqs.~(\ref{defcalJ}), (\ref{defcalA}). Under the above
conditions, we can slave $\rho_{1}$ to $\rho_{0}$ so that
\begin{equation}
(\gamma_{R}{\cal A}[1-n_{1}] + \gamma'_{R}{\cal A}[n_{1}])\rho_{1}
= (\gamma_{L}{\cal J}[1-n_{1}]+
	\gamma'_{L}{\cal J}[n_{1}])\rho_{0}
\end{equation}
Operating on both sides alternately by ${\cal J}[n_{1}]$ and ${\cal
J}[1-n_{1}]$ it is easy to show that
\begin{eqnarray}
	\gamma_{R}'{\cal J}[n_{1}]\rho_{1} & = & \gamma_{L}'{\cal
J}[n_{1}]\rho_{0} \\
	\gamma_{R}{\cal J}[1-n_{1}]\rho_{1} & = & \gamma_{L}{\cal
J}[1-n_{1}]\rho_{0} .
\end{eqnarray}
Substituting these into Eq.~(\ref{dotrho0}) yields
\begin{equation}
\dot{\rho}_{0} = (\gamma_{L}' + \gamma_{L}){\cal D}[n_{1}]\rho_{0}
\end{equation}

Since the probability or their being an electron on the SET is very
small we can say $\rho \simeq \rho_{0}$. Hence we have derived
The master equation (\ref{me1}) (without the Hamiltonian term)
for the dot alone
\begin{equation}
\dot{\rho} =  (2D_{0}+D_{1}){\cal D}[c_{1}^\dagger
c_{1}]\rho.
\end{equation}
Here we have defined
\begin{eqnarray}
D_{1} &=& \gamma_{L}^{\prime}-\gamma_{L},\\
D_{0} &=& \gamma_{L}.
\end{eqnarray}

Because the SET collector reservoir in the two cases (an electron on the
dot and an electron not on the dot) are independent (due to the SET
energy shift), the state conditioned on an electron entering the
collector is an incoherent mixture of the two possible paths. From
quantum trajectory theory \cite{Carmichael93,Wiseman93}, the
unnormalized state conditioned on this event is
\begin{equation}
dt\left(\gamma_{R}{\cal J}[b(1-n_{1})] + \gamma_{R}'{\cal
J}[bn_{1}]\right)R .
\end{equation}
The norm of this state matrix gives the probability for this event,
and is equal to the norm of
\begin{equation}
dt\left(\gamma_{R}{\cal J}[1-n_{1}] + \gamma_{R}'{\cal
J}[n_{1}]\right)\rho_{1} .
\end{equation}
>From the adiabatic elimination procedure above, this is equal to
\begin{equation}
dt\left(\gamma_{L}{\cal J}[1-n_{1}] + \gamma_{L}'{\cal
J}[n_{1}]\right)\rho .
\end{equation}
This is the unnormalized state $\tilde{\rho}_{1}(t+dt)$
of the dot alone conditioned on an
electron tunneling through the SET. From this it is easy to derive the
rate of such tunnelings as
\begin{equation}
\gamma_{L}\left\langle 1-n_{1}\right\rangle +
\gamma_{L}'\left\langle n_{1}\right\rangle =
D_{0}+D_{1}\left\langle{c_{1}^\dagger c_{1}}\right\rangle.
\end{equation}

%\end{multicols}

\begin{figure}[tbp]
\caption{Schematic representation of a single electron measurement for two
coupled quantum dots}
\label{fig1}
\end{figure}

\begin{figure}[tbp]
\caption{SET using coulomb blockade for a single electron measurement. The
Coulomb blockade gap is labeled $E_{CB}$ and the tunneling rates in the `on'
position are $\gamma_L$ and $\gamma_R$ through the left and right barriers
respectively. In (a) the electron is localized on dot-2 and a background
current $eD_0$ flows. In (b) the electron is localized on dot-1 and the
current $e(D_0+D_1)$ flows in the SET. }
\label{fig2}
\end{figure}

\begin{figure}[tbp]
\caption{A plot of the noise power spectrum normalized by the shot noise
level for $D_0=0$. (a) $\Omega=0.1$, (b) $\Omega=0.5$, (c) $\Omega=5.0$.}
\label{fig3}
\end{figure}

\begin{figure}[tbp]
\caption{A plot of the conditional population difference dynamics of $z_{%
{\rm c}}(t)$ versus scaled time for various values of the tunneling rate. In
all cases $D_0=0$ and time is measured in units of $D_1^{-1}$. (a) $\Omega=0.1$%
, (b) $\Omega=0.5$, (c) $\Omega=5.0$.}
\label{fig4}
\end{figure}

\begin{figure}[tbp]
\caption{A plot of the average of the conditional quantity $z_{{\rm c}}^2$
versus $\Omega/D_1$, wich measures the extent to which the measurement
localises the state of the dot. We set $D_0=0$. The solid lines refer to
the exact result Eq(59) }
\label{fig5}
\end{figure}

\begin{figure}[tbp]
\caption{The effect of finite background current with $D_0=1.0$ and $%
\Omega=1.0$. A plot of (a) the 'purity measure' $x_{{\rm c}}^2+y_{{\rm c}%
}^2+z_{{\rm c}}^2$ versus scaled time (units of $D_1^{-1}$) and (b) the
conditional population difference $z_{{\rm c}}(t)$ for a typical trajectory.
}
\label{fig6}
\end{figure}

\begin{figure}[tbp]
\caption{The effect of finite background current with $D_0=10.0$ and $%
\Omega=1.0$. A plot of (a) the 'purity measure' $x_{{\rm c}}^2+y_{{\rm c}%
}^2+z_{{\rm c}}^2$ versus scaled time (units of $D_1^{-1}$) and (b) the
conditional population difference $z_{{\rm c}}(t)$ for a typical trajectory.
}
\label{fig7}
\end{figure}


\begin{references}
\bibitem{Kane98}  B.E.Kane, Nature, {\bf 393}, 133 (1998).

\bibitem{Loss98}  D. Loss and D. P. DiVincenzo, Phys. Rev. A 57, 120 (1998).

\bibitem{Schoen97}  A.Shirman,G.Schoen, and Z. Hermon, Phys. Rev. Lett. 79,
2371 (1997).

\bibitem{Gurvitz97}  S. A. Gurvitz. Phys. Rev B, 15 215 {\bf 56},(1997)

\bibitem{Schoen98}  Alexander Shnirman and Gerd Schoen,{\em Quantum
Measurements Performed with a Single-Electron Transistor}, cond-mat/ 9801125
v3 12 Feb (1998).

\bibitem{Blick98}  , R.H.Blick, D.Pfannkuche,R.J. Haug, K.v.Klitzing and
K.Eberl, Phys. Rev. Letts, {\bf 80},4032 (1998).

\bibitem{WallsMilb94}  D.F. Walls and G.J. Milburn, Quantum Optics, pages
92-97, (Springer, Berlin 1994)

\bibitem{Sch99}  G. Sch\"on, A. Shnirman, and Y. Makhlin, cond-mat/9811029

\bibitem{Carmichael93}  H. Carmicael, An Open Systems Approach to Quantum
Optics, Springer- Verlag (1993).

\bibitem{DalCasMol92}  J. Dalibard, Y. Castin and K. M\o lmer,
``Wave-Function Approach to Dissipative Processes in Quantum Optics'' Phys.
Rev. Lett. {\bf 68}, 580 (1992).

\bibitem{Wiseman93}  H. M. Wiseman and G.J . Milburn, Phys. Rev. A {\bf 47}
, 1652 (1993) (appendix).

\bibitem{Sun99}  He Bi Sun and G.J. Milburn, "Quantum open systems approach
to current noise in resonant tunneling structures", Phys Rev B,{\bf 59}, 10748-10756, (1999).


\bibitem{Schoelkopf98}  R.J.Schoelkopf, P.Wahlgren, A.A.Kozhenikov,
P.Delsing and D.E.Prober, Science, {\bf 280} 1238, (1998).

\bibitem{Kane99}  B.E.Kane, N.S.McAlpine, A.S.Dzurak, R.G.Clark,
G.J.Milburn, He Bi Sun and H.M. Wiseman, Phys. Rev B {\bf 61}, 2961 , (1999) .

\bibitem{Joyez97}  P.Joyez, V.Bouchiat, D. Esteve, C. Urbina and
M.H.Devoret, Phys. Rev. Lett, {\bf 79}, 1349 (1997).

\bibitem{datta}  S.Datta,{\em Electronic transport in mesoscopic systems},
(Cambridge University Press, Cambridge, 1995).

\bibitem{milburn}  G.J.Milburn, {\it Quantum stochastic processes in
mesoscopic conductors}, cond-mat/99, submitted to J.Phys.A (1999).

\bibitem{Wis93a}  H.M. Wiseman,
%``Stochastic quantum dynamics of a continuously monitored laser''
Phys. Rev. A {\bf 47}, 5180 (1993).
\end{references}
\end{document}